\documentclass[12pt]{article}
\usepackage{amssymb}
\usepackage{amstext}
\usepackage{amsmath}
\usepackage{latexsym}
\usepackage{color}

\newcommand{\be}{\begin{equation}}
\newcommand{\en}{\end{equation}}

\newtheorem{thm}{Theorem}
\newtheorem{cor}[thm]{Corollary}

\newtheorem{defi}{Definition}[section]
\newtheorem{lem}[defi]{Lemma}
\newtheorem{theorem}[defi]{Theorem}

\newcommand{\bedefin}{\begin{defi}}
\newcommand{\findefi}{\end{defi} \medskip}

\newcommand{\betheo}{\begin{theorem}$\!\!${\bf \,\,\,}}
\newcommand{\entheo}{\end{theorem}}
\newcommand{\enth}{\end{theorem}}

\newcommand{\becor}{\begin{cor}$\!\!${\bf .}}
\newcommand{\encor}{\end{cor}}

\newcommand{\belem}{\begin{lem}$\!\!${\bf .}}
\newcommand{\enlem}{\end{lem}}

\newcommand{\bea}{\begin{eqnarray}}
\newcommand{\ena}{\end{eqnarray}}

\newcommand{\beano}{\begin{eqnarray*}}
\newcommand{\enano}{\end{eqnarray*}}

\newcommand{\bee}{\begin{enumerate}}
\newcommand{\ene}{\end{enumerate}}

\newcommand{\bei}{\begin{itemize}}
\newcommand{\eni}{\end{itemize}}

\newcommand{\betab}{\begin{tabular}}
\newcommand{\entab}{\end{tabular}}

\newcommand{\bd}{\begin{displaymath}}

\newcommand{\h}{{\mathfrak H}}

\newcommand{\hk}{{\mathfrak H}_{K}}
\newcommand{\htil}{\widetilde{\mathfrak H}}

\newcommand{\bPhi}{\mbox{\boldmath $\Phi$}}
\newcommand{\bPsi}{\mbox{\boldmath $\Psi$}}

\catcode `\@=11 \@addtoreset{equation}{section}

\catcode `\@=12

\begin{document}

\begin{center}

{\Large \bf Coherent States and Bayesian Duality}\vspace{0.5cm}\\

\vspace{1cm}

{\large S. Twareque Ali} \footnote[1]{Department of Mathematics and
Statistics, Concordia University,
Montr\'eal, Qu\'ebec, CANADA H3G 1M8\\
e-mail: stali@mathstat.concordia.ca}
\vspace{3mm}\\

{\large J.-P. Gazeau} \footnote[2]{Astroparticules et Cosmologie
(APC, UMR 7164),
Universit\'e Paris  Diderot Paris 7,
10, rue Alice Domon et L\'eonie Duquet,
75205 Paris Cedex 13\\
e-mail: gazeau@apc.univ-paris7.fr
}
\vspace{3mm}\\

{\large B. Heller} \footnote[3]{
Department of Applied Mathematics,
Illinois Institute of Technology,
Chicago, IL 60616\\
e-mail: heller@iit.edu, effe@midway.uchicago.edu
}
\end{center}

\begin{abstract}
\noindent
We demonstrate how large classes of discrete and continuous statistical  distributions can be
incorporated into coherent states, using the concept of a reproducing kernel Hilbert space.
Each family of coherent states is shown to contain, in a sort of duality, which
resembles an analogous duality in Bayesian statistics, a discrete
probability distribution and a discretely parametrized family of continuous distributions.
It turns out that nonlinear coherent states, of the type widely studied in quantum optics,
are a particularly useful class of coherent states from this point of view, in that they
contain many of the standard statistical distributions. We also look at vector
coherent states and multidimensional coherent states as carriers of
mixtures of probability distributions and
joint probability distributions.

\end{abstract}

\newpage

\tableofcontents


\newpage

\parskip=6pt

\section{Introduction}\label{intro}
In a series of recent papers, \cite{hellwang2007,hellwang2006,hellwang2004}, an
intimate connection between certain families of coherent states and statistical
distributions has been demonstrated and studied. The coherent states discussed in these
papers all have group theoretical origins and the Haar measure on the group has then
been shown to induce a prior measure on the statistical parameters entering
the definition of the discrete distributions. In this paper we look at a broader class of
coherent states, which do not necessarily have their origins in group representations.
In particular we show how, under certain technical restrictions, we can start with a
discrete probability distribution, depending on a single real parameter, and associate
coherent states to it. In the process we
obtain a natural family of discretely indexed continuous distributions, which are then
in a
sort of {\em duality} with the original discrete distribution, via the coherent states.
This duality is highly reminiscent of a similar
duality observed in the theory of Bayesian statistics, since the
resolution of the identity condition, which we impose on the coherent states,
introduces a preferred  {\em prior measure} on the parameter space
of the discrete distribution, with this distribution itself playing the role of the
{\em likelihood function\/.}
The associated discretely indexed continuous distributions become the related
{\em conditional posterior distributions\/.} Alternatively,
one can also start with a discretely parametrized family  of continuous distributions,
and
under a certain convergence assumption, once more build coherent states. These coherent
states then
again give rise to a dual discrete distribution or likelihood
function. We illustrate the theory by
looking
at a few examples of well-known statistical distributions
(additional examples may be found in \cite{gazconc}). Although most of these
examples
have been studied earlier, in the context of Glauber-Klauder-Sudarshan or
Gilmore-Perelomov coherent states
\cite{hellwang2007,hellwang2006,hellwang2004}, we analyze them here from the
present
perspective,
i.e., without invoking any group property.

  We take the discussion further by studying the relevance of vector coherent
states and multidimensional coherent states when mixtures of  probability distributions
or joint
distributions are considered. As far as we are aware, this is the first
time that such vector
coherent states have been studied in connection with statistical distributions.

\section{Experimental model context}

 In the following paragraphs, using simple
experimental setups, we try to motivate the simultaneous appearance of a family
of discrete probability distributions and a family of continuous distributions
in the sort of duality referred to earlier. First we describe a classical
statistical procedure known as Bayesian inference. Then, as indicated above,
we will consider a relationship between our subsequent mathematical
analysis and this classical procedure. (See Appendix.)

\subsection{Discrete Case}

Suppose we have an experimental setup for which we have an ``experimental
    model'' in the
form of a family of discrete probability distributions $n \mapsto
P(n,\lambda )$  relating to a discrete set of
possible experimental outcomes. That is, we
do not know the preparation exactly, only to the extent of a family
of states, indexed, say by the parameter $\lambda$ which takes
(continuous) values in some parameter space.  The parameter usually
represents a quantitative property of interest. In fact, the whole
idea of the experiment, presumably, is to obtain data with which to
estimate this physical property represented by the parameter.   As
an elementary example, let us think in terms of setting up an
experiment to toss a coin $N$ times and count the total number, $k$,
of heads. Now perform the experiment and designate the observed
value of $k$ as $k_{\text{obs}}$.  Then use $k_{\text{obs}}$ to
estimate the bias of the coin.  The statistical model would be a
family of binomial distributions indexed by a parameter $p$ with ``true''
but unknown parameter value $p_0$. One can estimate the value of
$p_0$ as $p_{\text{est}}=k_{\text{obs}}/N$.  But
conditionally upon
the observed value, $k_{\text{obs}}$, one may consider $p$ as a random variable and
construct a certain conditional probability distribution over the
parameter space which we now treat as a measurable space. The
motivation for this inference procedure is that, for example, one could
then find subsets of the parameter space for which one could make
statements such as ``given the result of the experiment,
there is a $99\%$ chance that the true value $p_0$ lies within that
subset''. (Think of an experiment where one tossed a coin $1000$
times and got $999$ heads.)

\subsection{The duality}\label{subsec-duality}
In the Bayesian context, both the quantity to be observed and the unknown
parameter are considered to be random quantities, playing a
dual role.   We consider two
conditional probability distributions. Before performing the random
experiment, the experimental model in the form of a family
$P(y, \lambda)$  of  discrete probability distributions is viewed
as a conditional distribution of the random variable $Y$  given the
parameter value, say $\lambda$. After performing the
experiment, we have an observed value, say $y_{\mathrm{obs}}$, and
we compute the conditional probability density function of the
parameter $\lambda$  given $y_{\mathrm{obs}}$, obtaining a posterior
conditional probability distribution. But, of course, we need to
choose a prior measure $P(d\lambda)$. Suppose we have a probability
density function where $P(d\lambda) = \Pi(\lambda) \,d\lambda$.The
posterior probability density function is then given by
\cite{boti,per} (see also the Appendix at the end):
\begin{equation}
f(\lambda , y_{\mathrm{obs}}) = \frac{P( y_{\mathrm{obs}} ,\lambda)\,
\Pi(\lambda)}{\int P( y_{\mathrm{obs}} , \lambda')\, \Pi(\lambda')\, d \lambda'}\; .
\label{bayes-post}
\end{equation}

 A prototype classical example of the binomial distribution is the coin tossing
experiment mentioned above and given in the Appendix.  In that classical context,
the posterior conditional probability density function for the parameter $p$ would be
obtained  according to (\ref{bayes-post}).

An example of a Bayesian approach involving the
binomial distribution in a
quantum context is given in \cite{per}. A thought experiment is described involving a
count of photons which are passed though a polarizer, a pinhole, and a calcite
crystal, eventually triggering a detector as $(+)$ or $(-)$. In that context, a posterior
distribution is obtained via   (\ref{bayes-post}) for the binomial parameter
$\theta$, the direction  of the polarizer.

 In \cite{per}, the family of probability distributions which we have called the
stochastic model for the experiment is designated as {\em predictive}. The
conditional probability distribution for the parameter that we have called
Bayesian posterior is there designated as {\em retrodictive}.

\section{A general setting for statistical distributions and coherent states}
\label{sec-gen-setting}
Let $\{X, \mu\}$ be a measure space. $X$ could, for example, be the
space of some statistical parameters or a larger space containing
such parameters. Consider the Hilbert space $\h = L^2(X, \mu)$ and suppose that
it contains a reproducing kernel subspace $\h_K$. This means that for any orthonormal basis,
$\{\Phi_k\}_{k=0}^N$ of $\h_K$, (where $N$ could be finite or infinite) the following is true:
\begin{enumerate}
\item $\sum_{k=0}^N\vert \Phi_k (x)\vert^2 < \infty$,  for almost all
     $x \in X$ and in fact, it is possible to define the functions $\Phi_k (x)$
     in a way so that this convergence condition holds everywhere.
\item The function
\be
K(x,y) = \sum_{k=0}^N \Phi_k (x)\overline{\Phi_k (y)}
\label{rep-ker-def}
\en
 defines a
     {\em reproducing kernel\/,} i.e.,  $K(x,y)$ satisfies the properties,
\bea
  K(x, y) & = & \overline{K(y,x)}\; , \qquad K(x,x ) > 0 , \;\; \text{for all}\; x\in X\; ;
    \nonumber \\
  \int_X K(x,z) K(z,y)\; d\mu (z) & = & K(x,y ) , \;\; \text{for all}\; x, y \in X\; .
\label{repkerdef}
\ena
It turns out that the kernel is independent of the orthonormal basis
chosen to represent it.
\end{enumerate}

For such a Hilbert space $\h_K$, we can define a set of vectors, $\vert x \rangle$, labelled by the
points of $X$ in the manner:
\be
  \vert x \rangle =  {\mathcal N} (x)^{-\frac 12}K(.\;, x ) =
 {\mathcal N} (x)^{-\frac 12}\sum_{k=0}^N \overline{\Phi_k (x)}\Phi_k\; , \qquad
{\mathcal N} (x) = K(x,x) = \sum_{k=0}^N\vert \Phi_k (x)\vert^2\; .
\label{CSdefin}
\en
The normalization factor ${\mathcal N} (x)$ is chosen in order to ensure that
$\langle x\mid x\rangle =1$.
In view of (\ref{repkerdef}), these vectors are then immediately seen to
satisfy the {\em resloution of the identity\/.}
\be
  \int_X \vert x \rangle\langle x \vert\;{\mathcal N} (x)\; d\mu (x) = I_{\h_K}\; ,
\label{resoliddef}
\en
This condition implies that the vectors $\vert x \rangle$ form an {\em overcomplete}
set in $\h_K$, so that any vector in it can be written as a linear combination,
either as a sum of or an integral over  these.
Very often such a set of vectors is associated to a unitary representation of some group,
and are constructed by letting  the representation operators act on a fixed vector in
$\h_K$. At other times such vectors are obtained by exploiting analytic properties
of  vectors in $\h_K$. But at this point,
we prefer to adopt a more general point of view and to just focus on the reproducing
kernel Hilbert space structure.  We shall call the vectors $\vert x \rangle$
{\em (generalized) coherent states\/,} (see, for example \cite{aagbook}, for a detailed
discussion).

  It is possible to associate two  types of probability distributions to the basis vectors
in a reproducing kernel Hilbert space. First, writing
\be
  P(n, x )=  \frac {\vert\Phi_n (x)\vert^2}{\mathcal N (x)}, \qquad n =0,1,2, \ldots , N\; ,
\label{prob-dist1}
\en
we see that $\sum_{n=0}^N P(n, x) = 1$. Thus, $P(n,x)$ can be looked upon as a
{\em discrete probability distribution} with parameter $x$. For instance, it can
be based upon some experimental setup and then might be viewed as a stochastic model.
Secondly, if $X \subset
\mathbb R^m$, and if $d\mu$ has a  Radon-Nikodym density with respect to the Lebesgue
measure $dx$ (on $\mathbb R^m$), then the functions,
\be
   \Psi_n (x ) = \vert\Phi_n (x)\vert^2\; \frac {d\mu (x)}{dx}
   = P(n, x)\;\mathcal N (x) \frac {d\mu (x)}{dx},
   \qquad n=0,1,2, \ldots , N\; ,
\label{prob-dist2}
\en
define, for each $n$ a {\em continuous probability density} on $X$,
since $\int_X \Psi_n (x)\; dx = 1$.  In the context of Bayesian
statistics, this could be thought of as a
conditional probability density for $x$, given $n$.  If $P(n,x)$ is a statistical distribution,
corresponding to some physical situation, which depends on the parameter $x$, the measure
\be
 d\overline{\kappa}(x) = \mathcal N (x)\;d\mu (x)
\label{post-distrib}
\en
can  be interpreted as a prior measure on the parameter space $X$
and then the
$\Psi_n (x)$ become the associated posterior distributions, in conformity with
(\ref{bayes-post}).
In \cite{hellwang2007,hellwang2006,hellwang2004},
a group theoretical argument, exploiting the invariant measure and coherent states
related to a  particular representation of the group on a Hilbert space, were invoked to
obtain the prior measure.  Here we see that the appearance of a discrete
probability distribution $P(n, x)$ and the continuous probability
distributions $\Psi_n (x)$ in this  dual relationship is embodied in the structure of
the coherent states  $\vert x \rangle$, independently of any group action.

\subsection{A generic example}\label{subsec-gen-ex}

  As a particular example, of the above situation, which will be useful for
the purposes of the present paper, and which will turn out to have rich applications
to statistical distributions encountered in extensive physical contexts, we
introduce a family of the so-called {\em non-linear coherent states\/.} These are
built by taking an abstract, complex, separable Hilbert space $\h$, of dimension
$N$ (finite or infinte), choosing an orthonomal basis
$\phi_k \;, k = 0 , 1, 2, \ldots , N$, of it and defining on it the vectors
\be
  \vert z \rangle = {\mathcal N}(\vert z\vert^2)^{-\frac 12}\; \sum_{k=0}^N
     \frac {z^k}{[x_k !]^{\frac 12}}\; \phi_k\; ,
\label{DCS}
\en
where $z$ is a parameter drawn from some appropriate open subset of $\mathbb C$ and
$x_1 , x_2 , x_3 , \ldots ,$ is a conveniently chosen positive sequence of numbers
for which we define the generalized factorial,
$ x_k ! =  x_1 x_2 \ldots x_k$, with $x_0 ! =1$, by definition. The normalization
factor in this case is ${\mathcal N}(\vert z\vert^2) =
\sum_{k=0}^N\dfrac {\vert z \vert^{2k}}{x_k !}$ and of course,
$\langle z \vert z \rangle = 1$. In order to ensure that these coherent states  form
an overcomplete set of vectors in the
Hilbert space $\h$, one requires the resolution of the identity,
\be
 \int_{\mathcal D}\vert z \rangle\langle z \vert\; {\mathcal N}(\vert z\vert^2)
 \;d\nu (z, \overline{z} )  = I_\h\; ,
 \label{resolid}
\en
to hold, where $I_\h$ is the identity operator on the Hilbert space
$\h$ and $\mathcal D$ is an appropriate domain of the
complex plane (usually the open unit disc or an open annulus, but which could also be
the entire plane).
It is not hard to see that the resolution of the identity (\ref{resolid})
 will hold if the measure $d\nu$, which  is usually
of the type $d\varrho (r)\;d\theta$ (for $z = re^{i\theta}$), is such that $d\varrho$
is related to the $x_k !$ through the following moment condition
(see, for example, \cite{sim97} for a discussion of the moment problem):
\be
  \frac {x_k !}{2\pi} = \int_0^{\sqrt{L}} r^{2k}\; d\varrho (r )\;  ,
\qquad k=0,1,2, \ldots ,
\label{momprob}
\en
$L$ being the radius of convergence of the series
$\sum_{k=0}^N\dfrac {\vert z\vert^{2k}}{x_k !}$ (considered as a series in $\lambda =
\vert z\vert^2$). This
means that once the sequence  $x_1,x_2, x_3, \ldots , $ is specified, the measure
$d\varrho$ is to be determined
by solving the moment problem (\ref{momprob}). There is an extensive literature on
the construction of  coherent states of this type (see, for example,
\cite{gazklau,klaupensix,manmarsuza,odz98}).
On the other hand, if the moment problem has no solution or, it has a solution but the
corresponding measure is not  explicitly known, there exists an alternative
constructive procedure which allows one to build non-linear coherent states, again
resolving the identity \cite{collcs}.

  We proceed now to analyze the discrete and continuous probability distributions, in
the sense of the previous section, associated to these coherent states.

\subsection{Discrete distribution associated to $\vert z \rangle$}
\label{subsec-disc-distrib}
With $\lambda = \vert z\vert^2$, define the discrete probability distribution
$P(n, \lambda ), \; n =0,1,2, \ldots , N$, by
\be
  P(n, \lambda ) = \frac {\lambda^n}{x_n!}\;
  \mathcal N (\lambda )^{-1}\; .
\label{CS-assoc-discr-distrib}
\en
The normalization condition $\langle z\mid z\rangle =1$ is seen to imply that
\be
 \sum_{n=0}^N P(n, \lambda ) =1\; .
\label{prob-normaliz}
\en

In the special case, where $x_n = n$, this distribution is just the well-known Poisson
distribution, for then $x_n! = n!, \; N (\lambda ) = e^\lambda$ and  $L = \infty$.
We shall see later that many of the  well-known discrete  statistical distributions are
related to nonlinear coherent states in this manner.
Note that if $Y$ denotes the discrete random variable, $Y(n) = x_n$, then taking
$x_0 = 0$, we obtain its expectation value,
\be
\langle Y\rangle = \sum_{n=0}^N x_n P(n, \lambda ) = \lambda\; .
\label{exp1}
\en
Thus for each $\lambda$ we get a discrete probability distribution, which is some sort of
a generalized Poisson distribution. In general, the sort of distributions given by
(\ref{CS-assoc-discr-distrib}) are of the power series type, well-known in statistics
(see, for example \cite{joko}).

\subsection{Continuous distributions associated to $\vert z \rangle$}
\label{subsec-conts-distrib}
We next note that in view of (\ref{momprob}),
$$
   2\pi\int_0^L P(n, \lambda )\; \mathcal N (\lambda )\;
   d\overline{\varrho}(\lambda) = 1, \qquad
   n = 0, 1, 2, \ldots , N,$$
where we have written
\be
 d\overline{\varrho}(\lambda) = d\varrho (r), \qquad r^2 = \lambda\; .
\label{var-change}
\en
Thus, the functions,
\be
  \Psi_n (\lambda ) = 2\pi P(n, \lambda )\; \mathcal N (\lambda )
  \; \frac {d\overline{\varrho} (\lambda)}{d\lambda}
   = 2\pi \frac {\lambda^n}{x_n!}\; \frac {d\overline{\varrho} (\lambda)}{d\lambda},
\qquad n = 0, 1,2, \ldots ,
\label{CS-assoc-conts-distrib}
\en
define, for each $n$, a  continuous probability density over the parameter space
$0\leq \lambda \leq L$. Here, $\dfrac {d\overline{\varrho}(\lambda)}{d\lambda}$ denotes the
Radon-Nikodym derivative of the measure $d\overline{\varrho}$ with respect to the Lebesgue
measure $d\lambda$, provided it exists. Clearly,
\be
  \int_0^L\Psi_n (\lambda )\; d\lambda = 1, \qquad n= 0, 1, 2, \ldots \; .
\label{conts-prob-norm}
\en

From (\ref{prob-normaliz}) it follows that
\be
   \sum_{n=0}^N \Psi_n (\lambda ) = 2\pi  \mathcal N (\lambda )
  \; \frac {d\overline{\varrho} (\lambda)}{d\lambda} < \infty\; ,
\label{repker-cond}
\en
for almost all $\lambda \in [0, L]$. Also, if $\Lambda$ is the continuous
random variable over the parameter space $[0, L]$, such that $\Lambda (\lambda)
= \lambda$, then
\be
 \langle \Lambda \rangle_n =  \int_0^L \lambda \Psi_n (\lambda )\;
 d\lambda = x_{n+1}\; ,
\label{exp2}
\en
which is a dual relation to (\ref{exp1}).

  Finally note, that in terms of the discrete and continuous  probability distributions
themselves, the coherent states (\ref{DCS}) may be written as
\bea
  \vert z \rangle & = & \sum_{n=0}^N
  \left[P(n, \lambda )\right]^{\frac 12} e^{-in\theta}
  \phi_n \nonumber\\
  & = &  \left[2\pi \mathcal N (\lambda ) \; \frac {d\overline{\varrho} (\lambda)}{d\lambda}
  \right]^{-\frac 12} \sum_{n=0}^N \left[\Psi_n (\lambda)\right]^{\frac 12} e^{-in\theta}
  \phi_n\;, \qquad z = \sqrt{\lambda}e^{-i\theta}\; ,
\label{back-constr}
\ena
and which satisfy the resolution of the identity,
\be
  \int_0^L\!\!\int_0^{2\pi}\vert z \rangle\langle z \vert\;
  \mathcal N (\lambda )\;d\overline{\varrho}(\lambda)\;d\theta
  = I_\h\; .
\label{resolid7}
\en

Comparing (\ref{bayes-post}) and
(\ref{CS-assoc-conts-distrib}) we see that the measure
\be
   d\overline{\kappa} (\lambda ) = 2\pi \mathcal N (\lambda)\; d\overline{\varrho}
   (\lambda ),
\label{generic-post}
\en
gives a  prior measure on the parameter space $[0, L]$.  Furthermore, these
results give us a hint as to how one might construct coherent states starting from
families of probability distributions.

We emphasize again that the  duality appearing here, between the family of discrete
probability distributions, $n\longmapsto P(n, \lambda )$, parametrized by
$\lambda$ and the family of continuous distributions $\lambda \mapsto
\Psi_n (\lambda )$, parametrized by $n$, is analogous to the
 {\em Bayesian duality\/}, that  we already referred to at the end of
Section \ref{subsec-duality}, between a discrete probabilistic model
$P (n, \lambda)$ and the continuous probability density function,
(see also the Appendix to this paper), and which is captured in the relation,
\be
  f (\lambda, n ) = \frac {P(n, \lambda)\Pi(\lambda)}
  {\int_0^\infty P(n, \lambda) \Pi(\lambda)\; d\lambda},
\label{bayes-duality}
\en
where $n$ represents an experimentally realized value of the discrete random variable
and this conditional density function ({\em Bayesian posterior} density function) is
obtained using the prior measure $\Pi(\lambda)d\lambda$
(see, for example, \cite{kojo}).

It is interesting to note that the coherent states
  $\vert z \rangle$,
which are unit vectors  in the
Hilbert space $\h$, may be thought of as being {\em square roots} of the discrete
probability distribution function $n \mapsto P(n, \lambda )$,
in the sense that $\Vert\!\;\vert z\rangle\!\; \Vert^2 = \sum_{n=0}^N P(n, \lambda ) = 1$.

  The probability distribution $P(n, \lambda)$ can be extracted from the coherent
state  $\vert z \rangle$ by taking the trace:
\be
  P(n, \lambda ) = \text{Tr}[\vert z \rangle\langle z \vert\; \mathbb P_n ] =
   \vert\langle\phi_n\mid z\rangle\vert^2 \; ,
\label{prob-extrac}
\en
where $\mathbb P_n = \vert\phi_n\rangle\langle \phi_n\vert$. In a quantum mechanical
interpretation, this $P(n, \lambda )$ is the probability of measuring the physical
quantity encoded by the state  $\phi_n$ when the system under observation
had been  prepared in the state $\vert z\rangle$.

\subsection{Coherent states from discrete statistical distributions}
\label{subsec-cs-from-disc-stdis}
Suppose now that we start with a discrete probability distribution, $P(n, \lambda)$, where again
$n = 0, 1, 2, \ldots , N$, with $N$ being either finite or infinite and $\lambda$ is a parameter
drawn from the interval $[a,b] \subset [0, \infty)$. Of course, $\sum_{n=0}^N P(n, \lambda ) =1$
and we further assume that $P(n, \lambda )$ satisfies the conditions:
\begin{enumerate}
\item There exists a measure $d\kappa$ on $[a,b]$, absolutely continuous
with respect to the Lebesgue measure $d\lambda$ and such that
\be
   \int_a^b P(n, \lambda)\; d\kappa (\lambda ) := c_n < \infty, \qquad n = 0,1,,2, \ldots,
   N\; .
\label{cond1}
\en

\item For all $\lambda \in [a,b]$,
\be
   \sum_{n=0}^N \frac {P(n, \lambda )}{c_n} < \infty\; .
\label{cond2}
\en
\end{enumerate}
On the interval $[a,b]$, let us define the functions
\be
   \Psi_n (\lambda )= \frac 1{c_n}\; P(n, \lambda )\;
       \frac {d\kappa(\lambda )}{d\lambda}\;,
\label{fcns1}
\en
for which we note that
\be
  \int_a^b \Psi_n (\lambda )\; d\lambda = 1\; , \qquad n =0,1,2, \ldots, N\; ,
\label{fcns1-norm}
\en
and using them we define on the open annulus,
\be
   \mathcal D = \{z = \sqrt{\lambda}\;e^{-i\theta}  \mid  a < \lambda < b\;, \;\;\;
   0\leq \theta < 2\pi \} \subset \mathbb C\;,
\label{annular-dom}
\en
the functions
\be
  \Phi_n (z) = \frac 1{\sqrt{2\pi}}\left[\Psi_n (\lambda )\right]^\frac 12 e^{-in\theta}\;.
\label{fcns2}
\en

Note that the range of values of the index $n$ need  not be constrained to lie among
the nonnegative integers only. It could also be a subset of $\mathbb Z$ or all of it.

It is worthwhile pointing out that the measure $d\kappa$ postulated in (\ref{cond1})
is not necessarily unique, which leaves the possibility of there being several such measures
which could be acceptable. In the case of the discrete distributions arising from
non-linear coherent states, the requirement of the resolution of the identity, i.e.,
the moment condition (\ref{momprob}) fixes the measure $d\kappa$.
Also the functions (\ref{fcns1}) are exactly like the $f(\lambda , n )$ in
(\ref{bayes-duality}), appearing in the duality studied in Bayesian statistics \cite{kojo,per}
although, unlike in that case, we have here the additional restriction (\ref{cond2}).

Clearly, the functions $\{\Phi_n\}_{n=0}^N$ form an orthonormal set:
\be
  \int_{\mathcal D}\overline{\Phi_m (z)}\;\Phi_n (z)\; d\lambda \; d\theta =
  \delta_{mn}\; .
\label{orthogonality5}
\en

  Let $\h$ denote the Hilbert subspace of $L^2 (\mathcal D , d\lambda\; d\theta )$
generated by these functions. Since,
\be
 \sum_{n=0}^N \vert \Phi_n (z)\vert^2 = \frac 1{2\pi}
  \frac {d\kappa(\lambda )}{d\lambda}\;
  \sum_{n=0}^N  \frac {P(n, \lambda )}{c_n}  < \infty \; ,
\label{repker-cond4}
\en
by virtue of (\ref{cond2}), $\h$ is a reproducing kernel Hilbert space. From the discussion
at the beginning of this
section (see (\ref{CSdefin})), we can then define coherent states in $\h$ as:
\be
  \vert z \rangle = \left[{\mathcal N (\lambda )}\right]^{-\frac 12}
    \sum_{n=0}^N \left[ \frac {P(n, \lambda )}{c_n}\right]^{\frac 12}e^{-in\theta} \Phi_n\; ,
    \qquad \mathcal N (\lambda ) =
     \sum_{n=0}^N \frac {P(n, \lambda )}{c_n}\; ,
\label{CS-from-disc-distr}
\en
which now satisfy the resolution of the identity
\be
  \frac 1{2\pi}\int_a^b\!\!\int_0^{2\pi}\vert z \rangle\langle z \vert\;\mathcal N (\lambda )\;
   d\kappa (\lambda)\;d\theta  = I_\h\; ,
\label{resolid9}
\en

Note that from (\ref{cond1}) and (\ref{fcns1}), we get
\be
  \Psi_n (\lambda) =  \frac {P(n, \lambda )\Pi (\lambda )}{\int_a^b P(n, \lambda )\Pi (\lambda )\; d\lambda}\; ,
  \quad \text{where,} \quad \Pi (\lambda ) = \frac {d\kappa (\lambda)}{d\lambda}\;,
\label{bayes-duality2}
\en
so that $d\kappa$ can be thought of (see (\ref{bayes-duality})) as a {\em prior measure}
on the parameter space $a < \lambda < b$ and the $\Psi_n$ as  the
associated Bayesian posteriors.

To make the connection with (\ref{CSdefin})
and (\ref{resoliddef}), we easily see that the coherent states (\ref{CS-from-disc-distr})
can also be written as
\be
  \vert z \rangle = \widetilde{\mathcal N}(\vert z \vert^2)^{-\frac 12}\sum_{n=0}^\infty
\overline{\Phi_n (z)}\;\Phi_n , \qquad \widetilde{\mathcal N}(\vert z \vert^2)
= \sum_{n=0}^\infty \vert\Phi_n (z)\vert^2\; ,
\label{CS-from-disc-distr2}
\en
and the resolution of the identity as
\be
  \frac 1{2\pi}\int_a^b\!\!\int_0^{2\pi}\vert z \rangle\langle z \vert\;
  \widetilde{\mathcal N} (\lambda )\;
   d\lambda\;d\theta  = I_\h\; ,
\label{resolid9-1}
\en

\subsection{Coherent states from continuous statistical distributions}
\label{subsec-cs-from-conts-stdis}
We now proceed to construct analogous families of coherent states from sets of
continuous probability distributions. Suppose that $\Psi_n (\lambda)\;, \;\; n=0,1,2,
\ldots ,N$, is a set of continuous probability densities defined over the set $I\subset
\mathbb R$. Evidently, they satisfy
$$
  \int_I\Psi_n (\lambda)\; d\lambda = 1, \qquad n=0,1,2, \ldots , N\; . $$
We assume in addition that
\be
  \widetilde{\mathcal N} (\lambda ) := \frac 1{2\pi}\sum_{n=0}^N \Psi_n (\lambda ) < \infty\; ,
  \qquad \lambda \in I\; ,
\label{conts-norm}
\en
Then, as before we construct the set of functions on $X = I\times [0, 2\pi )$:
\be
   \Phi_n (\lambda , \theta ) = \frac 1{\sqrt{2\pi}}[\Psi_n (\lambda )]^{\frac 12} e^{-in\theta},
\qquad n = 0,1,2, \ldots N\; ,
\label{fcns4}
\en
and note that they form an orthonormal set in $L^2 (X, d\lambda\; d\theta )$. Let $\h$
be the Hilbert subspace of $L^2 (X, d\lambda\; d\theta )$ generated by these vectors.
Then once again,
following (\ref{CSdefin}) we construct the coherent states in $\h$:
\be
  \vert \lambda , \theta \rangle =
  \widetilde{\mathcal N} (\lambda)^{-\frac 12}\sum_{n=0}^N \overline{\Phi_n (\lambda ,
  \theta )}\;\Phi_n\; ,
\label{CSdefin11}
\en
with $\widetilde{\mathcal N}(\lambda)$ as in (\ref{conts-norm}). These coherent states satisfy the
resolution of the identity,
\be
  \int_I\!\!\int_0^{2\pi}\vert \lambda , \theta \rangle\langle \lambda , \theta \vert\;
  \widetilde{\mathcal N} (\lambda )\;
   d\lambda\;d\theta  = I_\h\; ,
\label{resolid11}
\en

Clearly, the discrete distribution function this time is
\be
  P(n, \lambda ) = \frac {\Psi_n (\lambda )}{\widetilde{\mathcal N} (\lambda )} \; ,
\label{likelihood-fcn}
\en
with $\widetilde{\mathcal N} (\lambda )\; d\lambda$ the prior measure.

\section{Some illustrative examples}\label{sec-illus-ex}
In this section we construct coherent states for some standard statistical
distributions, following the general procedure outlined above. These coherent states
have been obtained before, using group theoretical arguments
\cite{hellwang2007,hellwang2006,hellwang2004} and we shall
indicate, in each case, the group theoretic relevance of the coherent states. Moreover,
in each case the interplay between the dual system of discrete and continuous
distributions, embodied in the coherent states will be explicitly demonstrated.

\subsection{Coherent states from the Poisson distribution}\label{subsec-poiss-cs}

   For the Poisson distribution,  the probability of $n$ successes, given that the average
number of successes is $\lambda > 0$, is
\be
  P(n, \lambda ) = \frac {e^{-\lambda}\; \lambda^n}{n!} \qquad \text{and}
  \qquad \sum_{n=0}^\infty P(n, \lambda ) =1 \; .
\label{poss-prob}
\en
Once again we would like to relate these to a family of coherent states. Also,
thinking of $\lambda$ itself as
a random variable, we would like to obtain a distribution function for it.
We start by introducing  the complex variable, $z = \sqrt{\lambda}\;e^{-i\theta}$,
and since $\int_0^\infty P(n, \lambda )\; d\lambda = 1$ for all $n$, we define
the functions (see (\ref{fcns2}))
\be
  \Phi_n (z) = \frac 1{\sqrt{2\pi}}\;\left[P(n, \lambda )\right]^{\frac 12} e^{-in\theta} =
 \frac 1{\sqrt{2\pi}}\;\left[\frac {\lambda^n e^{-\lambda}}{n!}\right]^{\frac 12}
 e^{-in\theta} \; , \qquad n=0,1,2, \ldots , \infty\; .
\label{fcns22}
\en
These functions are clearly orthonormal with respect to the measure $d\lambda\; d\theta$:
$$
  \int_0^\infty\!\!\int_0^{2\pi} \overline{\Phi_m (z)}\; \Phi_n (z)\; d\lambda\; d\theta
    = \delta_{mn}\;. $$
Let $\h \subset L^2(\mathbb C, d\lambda\; d\theta )$ be the (infinite dimensional separable)
Hilbert space generated by them.  Next we see that
conditions (\ref{cond1}) and
(\ref{cond2}) are satisfied with $d\kappa = d\lambda$ and $c_n = 1$ for all $n$.
Thus, following (\ref{CS-from-disc-distr}) we may
define coherent states on $\h$ as
\be
  \vert z \rangle = \sum_{n=0}^\infty \sqrt{P(n, \lambda )}\; e^{-in\theta}\;\Phi_n
   = e^{-\frac{\vert z \vert^2}2}\sum_{n=0}^\infty \frac {z^n}{\sqrt{n!}}\; \Phi_n,
\label{poiss_CS1}
\en
so that,
$$ \langle z \vert z \rangle =\sum_{n=0}^\infty P(n, \lambda ) = 1\; .$$
Again, the coherent states $\vert z \rangle$,
may be thought of as being square roots of the discrete
Poisson distribution, $n \mapsto P(n, \lambda )$.

   These coherent states also satisfy a resolution of the identity:
\be
  \frac 1{2\pi}\int_0^\infty\!\!\int_0^{2\pi}\vert z \rangle\langle z \vert\;
   d\lambda \; d\theta  = I_\h \; .
\label{poiss-resolid}
\en

It is clear that this time the prior measure on the parameter space
$0\leq \lambda < \infty$ is just the uniform distribution $d\lambda$,
with the
Bayesian posteriors being given by $\Psi_n (\lambda ) = P(n, \lambda )$.
The coherent states (\ref{poiss_CS1}) are the canonical
coherent states, well known in the physical literature (see, e.g., \cite{aagbook}).
Moreover, these coherent states are associated to a unitary representation of the
Weyl-Heisenberg group and the prior measure $d\lambda$ is  also obtainable
from the Haar measure of this group \cite{hellwang2004}.

  Finally, it ought to be pointed out that the continuous distribution given by the
function $\Psi_n (\lambda) = P(n, \lambda)$ is just a $\gamma$-distribution, for each $n$.
In other words,
the discrete Poisson distribution and the continuous $\gamma$-distributions
(which may now be thought of as being conditional
distributions for the
average number of success $\lambda$, given  $n$ successes) are in duality
through the canonical coherent states. Moreover, had we started with the $\gamma$-distribution
functions, $\gamma_n (\lambda ) = \dfrac {\lambda^{n-1} e^{-\lambda}}{\Gamma (n)}$, defined
$\Psi_n  = \gamma_{n+1}\; , \;\; n= 0,1,2, \ldots , \infty$ and followed
through the steps in Section \ref{subsec-cs-from-conts-stdis}, we would have arrived at
the same coherent states (\ref{poiss_CS1}).  In the field of statistics, the gamma
distribution is said to
be a {\em natural conjugate} to the Poisson sampling process \cite{kojo2}.

\subsection{Coherent states from the binomial distribution}\label{subsec-binom-cs}

   Consider the binomial distribution for $N$ independent trials, each having a probability
of success
$p$ and of failure $q= 1-p$. The probability of getting $n$ successes in these $N$ trials
is
\be
  P(n, p) =  \binom{N}{n}\, p^n q^{N-n} = \frac {N!}{(N-n)! n!} p^n q^{N-n} ,
  \qquad n= 0, 1, 2, \ldots  , N \; ,
\label{binom-prob1}
\en
and of course,
$$ \sum_{n=0}^N P(n, p) = (q + p)^N = 1 \; .$$
As before, we treat the parameter $p$  itself also as a random variable and then use our
general construction in order to: (1) obtain coherent states representing this distribution and
(2) find a posterior distribution for $p$. This case has also been worked out in
\cite{hellwang2006}, using coherent states of the rotation group and we shall
indicate the connection to this approach in the sequel.
Let us  first introduce a new parameter
$\lambda$, which will be more convenient for our purposes:
\be
   \lambda = \frac pq \quad \Longrightarrow \quad q = \frac 1{1+ \lambda}\;
\quad \text{and} \quad 0\leq \lambda < \infty.
 \label{new-param}
\en
Using this we introduce the complex variable $z = \sqrt{\lambda} e^{-i\theta}$ and note that
in terms of $\lambda$, the probability distribution (\ref{binom-prob1}) can be rewritten as
\be
  P(n, \lambda) = \frac {N!}{(N-n)! n!} \cdot\frac {\lambda^n}{(1+ \lambda)^N} =
                 \frac {\Gamma (N+1)}{\Gamma (N-n +1) \Gamma (n+1)}\cdot
     \frac {\vert z\vert^{2n}}{(1+ \vert z\vert^2)^N}
\label{binom-prob2}
\en

Since
$$
   (N+1)\int_0^\infty \frac {\lambda^n}{(1+\lambda)^{N+2}}\; d\lambda
  = (N+1)\int_0^1 q^{N-n}(1-q)^n\; dq =   \frac {n! (N-n)!}{N!}\; , $$
we take (see (\ref{cond1}))
\be
  d\kappa =  \frac {(N+1)}{(1+\lambda)^2}\; d\lambda\;,
  \qquad c_n = 1\; .
\label{binon-poster}
\en
Since $N$ is finite, (\ref{cond2}) is trivially satisfied. Thus, we take
\be
  \Psi_n (\lambda ) = P(n, \lambda )\frac {d\kappa (\lambda)}{d\lambda}
  = \frac {(N+1)!}{(N-n)! n!} \cdot\frac {\lambda^n}{(1+ \lambda)^{N+2}}
    \label{binom-cont-distr}
\en
and
\be
     \Phi_n (z) = \left[\frac {(N+1)!}{2\pi (N-n)! n!}\right]^\frac 12
    \frac {z^n}{(1+ \vert z\vert^2)^{\frac N2 +1 }}\;,  \qquad z\in \mathbb C\; , \quad
    n=0,1,2, \ldots , N\; .
\en
Clearly, these vectors are orthonormal:
$$
  \int_0^\infty\!\!\int_0^{2\pi} \overline{\Phi_m (z)}\Phi_n (z)\; d\lambda \; d\theta
   = \delta_{mn} $$
and we denote by $\h$ the $(N+1)$-dimensional Hilbert space generated by these vectors.
On this space we then have the coherent states,
\be
  \vert z \rangle = \sqrt{P(n,\lambda)}\; e^{-in\theta}\Phi_n
= \frac 1{(1 + \vert z \vert^2)^{\frac N2}} \sum_{n=0}^N
\frac {\sqrt{\Gamma (N+1)}\;z^n}{\sqrt{\Gamma (N-n +1) \Gamma (n+1)}}\;\Phi_n ,
\label{binom-CS}
\en
Note again, that since
$$ \langle z \mid z\rangle = 1 = \sum_{n=0}^N P(n, \lambda)  \; ,$$
for each $\lambda = \vert z \vert^2$, the coherent state $\vert z\rangle$ is sort of
a vectorial square
root of the probability  distribution $P(n, \lambda), \; n=0,1,2, \ldots , N$.
These coherent states satisfy the resolution of the identity,
\be
  \frac 1{2\pi}\int_0^\infty\!\!\int_0^{2\pi} \vert z \rangle \langle z \vert\;
d\kappa (\lambda )\;d\theta  =  \frac {N+1}{2\pi}\int_0^\infty\!\!\int_0^{2\pi}
\vert z \rangle \langle z \vert\;
\frac {d\lambda\;d\theta}{(1+ \lambda)^2}
= I_\h \; .
\quad
\label{binom-resolid}
\en

Next, introducing the new labels $N = 2j, \; k = n-j$, we write
\be
  \vert z \rangle =  (1+ \vert z \vert^2)^{-\frac N2}\sum_{k= -j}^j
\frac {\sqrt{\Gamma (2j + 1)}\;z^{k+j}}{\sqrt{\Gamma (j - k +1) \Gamma (j+k+1)}}\;\Phi_k \; ,
\label{binom_CS2}
\en
which are immediately recognized as being the Gilmore-Perelomov-Radcliffe type coherent states
\cite{acgt,radc,perel,aagbook} for the $(2j+1)$-representation
of $SU(2)$. Indeed, the vectors $\vert z \rangle$ may be rewritten in terms of the $SU(2)$
generators
$J_\pm , J_3$  and the lowest basis vector $\Phi_{-j}$ as:
\be
   \vert z \rangle = e^{zJ_+ }\;e^{\eta J_3}\;e^{-\overline{z}J_-}\Phi_{-j} =
   e^{\xi J_+ - \overline{\xi}J_-}\psi_{-j} := D(\xi)\Phi_{-j}\; ,
\label{binom_CS3}
\en
where, writing $z = -\tan\frac{\vartheta}2\; e^{-i\gamma}$,
$$ \xi = i\frac {\vartheta}2 \;e^{i\gamma} \qquad \text{and}\qquad  \eta = \log (1+ \vert z\vert^2)
   = 2\log \sec \frac \vartheta 2\; .$$

   Finally, note that by virtue of (\ref{binom-prob2}) and (\ref{binom-cont-distr}),
the measure
\be
 d\kappa (\lambda) = \frac {N+1}{(1 +\lambda)^2}\;d\lambda \qquad \text{or equivalently,}
 \qquad
 d\kappa (p) = (N+1)\; dp,
\label{lambda-meas2}
\en
gives in this case the prior measure (again uniform) of the parameter
$p$ over the interval $[0,1]$.

  Once again, it is clear that had we started  with the continuous distributions
(\ref{binom-cont-distr}), which are $\beta$-distributions of the first kind,
and followed through with the procedure in
Section \ref{subsec-cs-from-conts-stdis}, we would also have arrived at the coherent states
(\ref{binom-CS}). Thus, the continuous $\beta$-distributions of the first kind and the
discrete binomial distribution (statistical {\em conjugate pair}) are in duality
through the $SU(2)$ coherent states.

\subsection{Coherent states from the negative binomial and\\
 $\beta$-distributions}
\label{subsec-neg-bin-cs}
The negative binomial and the $\beta$-distributions have a dual relationship through
the coherent states arising from the discrete series representations of the
 $SU(1,1)$ group.  Recall that, for a fixed integer
 $m \geq  1$, the negative binomial
 distribution is given by,
 \be
   P(m,n; \lambda) = \frac {\Gamma (m+n)}{\Gamma (n+1) \Gamma (m)}\;
   \lambda^m(1-\lambda)^n\; , \qquad n =0,1,2, \ldots , \infty\; ,
\label{neg-bin-distrib}
\en
where the parameter $\lambda$ lies in the interval $(0,1)$.
The quantity $P(m,n,\lambda )$ can be thought of as being the probability that
$m+n$ is the number of  independent  trials that are necessary to obtain the
result of $m$ successes (the $(m+n)$-th trial being a success) when
$\lambda$ is the probability of success in a single trial.
The term negative binomial stems from the fact that
$$
  (1 - \lambda )^{-k} = \sum_{n=0}^\infty \frac {\Gamma (k+n)}{\Gamma (n+1) \Gamma (k)}\;
   \lambda^n\;, $$
from which it also follows that
\be
 \sum_{n=0}^\infty P(m,n; \lambda ) = 1\; .
\label{neg-binom-norm}
\en
The $\beta$-distribution is a continuous distribution, in the variable $\lambda \in
[0,1]$, with discrete parameters $m, n = 1,2,3, \ldots , \infty$,
\be
  \beta (\lambda ; m,n ) = \frac 1{B(m,n)}\lambda^{m-1}(1-\lambda )^{n-1}\; , \qquad
\int_0^1 \beta (\lambda ; m, n )\; d\lambda = 1\; ,
\label{beta-distrib}
\en
where,
$$\quad B(m,n) = \frac {\Gamma(m)\Gamma (n)}{\Gamma (m+n)} =
   \int_0^1 t^{m-1} (1-t)^{n-1}\; dt\; .$$

 We note that,
\be
  \beta (\lambda ; m+1, n + 1) = \frac {P(m, n ; \lambda )}{c_{m,n}}, \quad \text{with}
  \quad  c_{m,n} = \frac m{(m+n+1)(m+n)} \; ,
  \label{beta-neg-binom}
  \en
implying, by virtue of  (\ref{beta-distrib}),
\be
  \int_0^1 P (m,n; \lambda ) \; d\lambda  = c_{m,n} \qquad \text{and} \qquad
  d\kappa (\lambda ) = d\lambda \; .
\label{neg-bin-kappa}
\en
Thus,   (\ref{cond1}) is satisfied, with $c_n = c_{m,n}$ and  (\ref{cond2}) is
also satisfied since,
\be
\sum_{n=0}^\infty  \frac {P(m, n ; \lambda )}{c_{m,n}} = \frac {m+1}{\lambda^2}
\sum_{n=0}^\infty P(m+2, n; \lambda ) =  \frac {m+1}{\lambda^2} < \infty\;,
\label{cond2-satisf}
\en
by virtue of (\ref{neg-binom-norm}).

   Thus, for fixed $m\geq 1$ and $n=0,1,2,
\ldots , \infty$, we define, using  (\ref{fcns1}) and
 (\ref{beta-neg-binom}), the continuous distributions,
\be
  \Psi_{m,n} (\lambda ) =\frac {P(m,n ; \lambda )}{c_{m,n}}\;
   \frac {d\kappa (\lambda )}{d\lambda}
    = \beta (\lambda ; m+1, n+1) = \frac 1{B(m+1, n+1)}
      \lambda^m (1-\lambda )^n \; ,
\label{neg-bin-conts-dis}
\en
and the associated functions in the complex variable $\zeta =
\sqrt{\lambda}\;e^{-in\theta}\; ,
\;\; 0\leq \lambda < \infty \; , \;\; 0\leq \theta < 2\pi$,
$$
\Phi_{m,n} (\zeta) = \frac 1{\sqrt{2\pi}}\left[\Psi_{m,n} (\lambda )\right]^\frac 12
   e^{-in\theta}\; , $$
which satisfy the orthonormality condition
$$
   \int_0^\infty\!\!\int_0^{2\pi} \overline{\Phi_{m,n}(\zeta)}\; \Phi_{m,k} (\zeta)\;
     d\lambda\; d\theta = \delta_{nk}\; . $$
Denoting by $\h$ the (infinite dimensional separable) Hilbert space
spanned by these vectors, and noting that by (\ref{cond2-satisf}),
$$\mathcal N (\lambda ) = \sum_{n=0}^\infty \frac {P(m,n; \lambda )}{c_{m,n}} =
\frac {m+1}{\lambda^2} \; , $$
we define the coherent states
associated to the discrete negative binomial and continuous
$\beta$-distributions, on this space using (\ref{CS-from-disc-distr}):
\bea
  \vert\zeta; m\rangle & = & \mathcal N (\lambda)^{-\frac 12}\sum_{n=0}^\infty
  \left[\frac{P(m,n; \lambda )}{c_{m,n}}
\right ]^{\frac 12}
   e^{-in\theta}\Phi_{m, n}\nonumber\\
    &  = & \sum_{n=0}^\infty \left[
     \frac {\Gamma (m+n+2)}{\Gamma (m+2)
  \Gamma (n+1)}\right]^\frac 12 \lambda^{\frac m2 +1}(1-\lambda)^\frac n2
  e^{-in\theta}\;\Phi_{m,n}\; .
\label{neg-binom-cs}
\ena
These satisfy the resolution of the identity,
\be
    \frac {m+1}{2\pi}\int_0^1\!\!\int_0^{2\pi}\vert\zeta ,
     m\rangle\langle\zeta , m \vert\;
    \frac {d\lambda\;d\theta}{\lambda^2} = I_\h\; .
\label{neg-bin-resolid}
\en
while from (\ref{bayes-duality}), (\ref{neg-bin-kappa})
and (\ref{neg-bin-conts-dis}) we obtain the prior measure on the parameter
space $[0,1]$:
\be
    d\kappa (\lambda ) =  d\lambda\; .
\label{neg-bin-post}
\en
 Note that this measure is different from the one
obtained in  \cite{hellwang2007}, which was derived using a group theoretical
argument.  However, in the present case, $m =1,2, 3,  \ldots$, while in
\cite{hellwang2007} the value $m=1$ was excluded. The associated Bayesian posteriors this
time are the $\Psi_{m,n}\; , \;\; n= 0,1,2, \ldots , \infty$\; .

  Once again it is clear that if we start with the continuous $\beta$-distributions
(\ref{beta-distrib}), and construct coherent states following Section
\ref{subsec-cs-from-conts-stdis}, with $\Psi_n (\lambda ) = \beta (\lambda ; m +1,n+1)$,
we arrive at these same coherent states.

To make contact with the coherent states of the $SU(1,1)$ group let us introduce
the new complex variable $z = (1-\vert\zeta\vert^2)^{\frac 12}e^{-i\theta}=
(1-\lambda )^{\frac 12}e^{-i\theta}$ and write $m + 2 = 2j$. Then
in terms of this variable we get the coherent states
\be
\vert z; j\rangle =  (1-\vert z\vert^2)^j
\sum_{n=0}^\infty \left[ \frac {\Gamma (2j+n)}{\Gamma (2j)
  \Gamma (n+1)}\right]^\frac 12 z^n\; \Phi_{2j,\;n}\; , \qquad j =  \frac 32 ,
  2, \frac 52, \ldots \;.
\label{suII-cs}
\en
These are the Gilmore-Perelomov type coherent states arising from the
discrete series
representations \cite{aagbook,hellwang2007,perel} of $SU(1,1)$.
  Since we are assuming that $m\geq 1$, the
  representation corresponding to $j =1$ does not appear here. We
  observe that in the
mathematical literature, these coherent states are usually written without the factor of
$(1-\vert z\vert^2)^j$ appearing before the sum on the right
hand side of  (\ref{suII-cs}). This is because, unlike in our case, the Hilbert space for
the discrete series representations of  $SU(1,1)$ is taken to be the one consisting of all
holomorphic functions on the open unit disc of $\mathbb C$, which are
square-integrable with respect to the measure
$\dfrac{(2j - 1)}\pi\;(1-\vert z\vert^2)^{2j-2}\; dx\;dy$, where $z = x + iy$, and
the factor is absorbed into the measure.

  Note finally, that all three examples discussed here lead to coherent states of the
  non-linear type (see (\ref{DCS})). To summarize, we have seen that the canonical
coherent states combine in duality the continuous
$\gamma$-distributions  with the Poisson distribution, the coherent
states of the $SU(2)$ group so combine the continuous $\beta$-distributions
of the first kind with the
discrete binomial distribution and the coherent states obtained from the discrete series
representations of the $SU(1,1)$ group combine in duality the continuous $\beta$-distributions
with the discrete negative binomial distribution.

\section{Vector and multidimensional coherent states \\ from probability distributions}
\label{vcs-prob-dist}
So far we have considered only single discrete probability distributions and constructed
coherent states from them. We now look at a situation where several
independently distributed
random variables are at play. It will turn out that the appropriate type of coherent states
to associate to such situations are vector coherent states (VCS) of the type discussed in
\cite{aeg,thirali} or multidimensional coherent states of the type studied in \cite{novgaz}.

     Let us take a discrete probability distribution $P(n, \lambda )\; , \;\; n =0,
1,2, \ldots , N$ (finite or infinte). This is the probability distribution of the
discrete random variable $\mathfrak N$ such that $\mathfrak N (n) = n$ and assume
that it is of the type (\ref{CS-assoc-discr-distrib}), i.e, the associated
coherent states are of the non-linear type.  Assume now that
we have $M$ such independent, random variables, distributed with parameters $\lambda_1, \lambda_2,
\ldots , \lambda_M$, respectively, each drawn from the interval $[0, L]$. Then
\be
 P(\lambda_1, \lambda_2, \ldots , \lambda_M; n) = \frac 1M \sum_{i=1}^M P(n, \lambda_i)
\label{comp-prob}
\en
is the probability of $n$ ``successes'' coming from any one of these processes when
we are indifferent to which one it comes from. We now ask if there is a natural set of coherent
states that could incorporate such a system of distributions, along the lines of what we
saw earlier. It will turn out that a Hilbert space over a matrix domain, consisting of
normal matrices, will be appropriate for the construction of such coherent sates.   Recall
that a normal matrix $\mathfrak Z$ is defined by the condition $\mathfrak Z^*\mathfrak Z
= \mathfrak Z\mathfrak Z^*$  and if $\mathfrak Z$ is an $M\times M$ matrix,
it can be diagonalized by means of a unitary matrix, i.e.,
\be
   \mathfrak Z = U\; \text{diag}\;[z_1, z_2, \ldots , z_M ]\;U^*
\label{normal-mat}
\en
where, $U\in U(M)$ and the elements $z_i, \;\; i = 1,2,3, \ldots , M$, of the diagonal
matrix are complex numbers. Writing $z_i = \sqrt{\lambda}_i\; e^{-i\theta_i}$,
let $\Omega$ denote the set of all such matrices for which
$0 \leq \lambda_i < L\; , \;\; i=1,2,3, \ldots , M$.
We next define the matrix valued functions on the domain $\Omega$,
\be
  \bPhi_n (\mathfrak Z ) = \frac {\mathfrak Z^n}{\sqrt{x_n!}}\; , \qquad n=0,1,2, \ldots ,
N\; ,
\label{matrix-basis}
\en
and on $\Omega$ we  define the measure,
\be
 d\Omega (\mathfrak  Z, \mathfrak Z^* ) = dU\;\prod_{i=1}^M
     d\overline{\varrho}(\lambda_i )\; d\theta_i\;, \qquad
     \int_\Omega d\Omega (\mathfrak Z , \mathfrak Z^* ) = 1\; .
\label{matrix-meas}
\en
where $dU$ is the (normalized) invariant measure of $U(M)$ and $d\overline{\varrho}$ is the
measure introduced in (\ref{momprob}) and (\ref{var-change}).

  It  then follows that the functions $\bPhi_n$ satisfy the matrix orthogonality condition:
\be
  \int_\Omega \mathfrak Z^m \;\mathfrak Z^{*n}\;d\Omega (\mathfrak Z , \mathfrak Z^* )
    = \mathbb I_M \;\delta_{mn}\; ,
\label{matrix-orthog}
\en
where $\mathbb I_M$ is the $M\times M$ identity matrix. Let $\{\chi^i\}_{i=1}^M$ be an
orthonormal basis of $\mathbb C^M$ and define the $\mathbb C^M$-valued functions,
\be
   \bPhi_n^i (\mathfrak Z^*) = \bPhi_n (\mathfrak Z^* )\chi^i \; .
\label{bas-vect}
\en

     Note  that
$$\text{Tr}[\mathfrak Z \mathfrak Z^*] = \sum_{i =1}^M \vert z_{i} \vert^2 \; . $$
Also, the series,
\be
   \sum_{n=0}^\infty\text{Tr}[\bPhi_n (\mathfrak Z)^*\;\bPhi_n(\mathfrak Z)]
   = \sum_{n =0}^N\sum_{i=1}^M \bPhi_n^i(\mathfrak Z)^\dag\;\bPhi_n^i(\mathfrak Z)
   =\sum_{n =0}^N\sum_{i=1}^M
   \frac {\lambda_i^n}{x_n!}
\label{assoc-ser}
\end{equation}
converges for all $\lambda_i \in [0,L)$, which following the discussion at the beginning
of Section \ref{sec-gen-setting}, is the condition for building reproducing kernel
Hilbert spaces, which we now proceed to do.

Consider the Hilbert space $\htil = L^2_{\mathbb C^N} (\Omega , d\Omega )$ of
square-integrable,
$M$-component vector-valued functions on $\Omega$. The
vectors $\bPhi^i_k , \;\;
i=1,2, \ldots , M,\;\; k = 0, 1,2, \ldots , N$ are elements of this Hilbert space and
in fact, by virtue of  (\ref{matrix-orthog}), they form an orthonormal set in it:
$$ \langle \bPhi _m^i \vert \bPhi _n^j \rangle=
\int_\Omega \bPhi _m^i(\mathfrak Z)^\dag \bPhi _n^j (\mathfrak Z )
\; d\Omega (\mathfrak Z , \mathfrak Z^* ) =  \delta_{mn}\; \delta_{ij}\; . $$
Denote by $\hk$ the Hilbert subspace of
$\htil$ generated by this set of vectors. Then, in view of the convergence of
the series  in (\ref{assoc-ser}),
$$
   \sum_{i,k} \Vert \bPsi ^i_k (\mathfrak Z^* )\Vert^2 < \infty\; ,
   \qquad \forall \;\mathfrak Z^* \in \Omega\; . $$
Thus, $\hk$ is a reproducing kernel Hilbert space of analytic functions in the variable
$\mathfrak Z^*$,
with matrix valued kernel $K : \Omega \times \Omega \longmapsto C^{N\times N}$, given by
(see (\ref{rep-ker-def}))
\bea
  K(\mathfrak Z^{* \prime} , \mathfrak Z ) & = & \sum_{i, k}\bPhi ^i_k (\mathfrak Z^{\prime})
   \bPhi ^i_k (\mathfrak Z^* )^\dag =  \sum_{i, k} \frac {\mathfrak Z^{* \prime k}\chi^i
     \chi^{i\dagger} \mathfrak Z^k}{x_k!}\nonumber\\
     & = &\sum_{i, k}\frac {\mathfrak Z^{* \prime k}\;\mathfrak Z^k}{x_k!} \; ,
\label{reker1}
\ena
When $M=1, \; \mathfrak Z = z, \; \Omega = \mathbb C$ and $x_k! = k!$,
we get the well-known Bargmann kernel,
$$ K(\overline{z}^\prime , z ) = e^{\overline{z}' z}\; , $$
and $\hk$ is the Hilbert space of entire analytic functions in the variable $\overline{z}$.
This is the kernel associated to the canonical coherent states (\ref{poiss_CS1}).

   The vector coherent states associated to the reproducing kernel $K$ are
(see (\ref{CSdefin})) the vectors
$\vert \mathfrak Z ; i\rangle
   \in \hk\;$,
\be
  \vert \mathfrak Z ; i\rangle (\mathfrak Z^{*\prime }) =
  \mathcal N(\mathfrak Z^* , \mathfrak Z )^{-\frac 12} K(\mathfrak Z^{*\prime} ,
  \mathfrak Z ) \chi^i\;,
    \qquad \mathcal N(\mathfrak Z^* , \mathfrak Z ) =
    \frac {K(\mathfrak Z , \mathfrak Z^*)}{M}
\label{VCS1}
\end{equation}
defined for each $\mathfrak Z \in \Omega$ and $i=1,2, \ldots , M$.
Note that since $K(\mathfrak Z^* , \mathfrak Z )$ is a strictly positive-definite matrix:
\be
  K(\mathfrak Z^* , \mathfrak Z ) = U\;\text{diag}\;[\mathcal N (\lambda_1),
  \mathcal N (\lambda_2),\ldots, \mathcal N (\lambda_M)]\; U^*\; ,
\label{matrix-normaliz}
\en
where for each $i$, $\mathcal N (\lambda_i )= \sum_{k=0}^N = \dfrac {\lambda_i^k}{x_k!}$ is the same
normalization factor as in (\ref{DCS}),
the negative square root makes sense. The vector coherent states
(\ref{VCS1}) satisfy the resolution of the identity (compare with (\ref{resolid7})),
\be
  \sum_{i=1}^M \int_\Omega \vert \mathfrak Z; i\rangle\langle \mathfrak Z; i\vert\;
  \mathcal N (\mathfrak Z^* , \mathfrak Z )\;d\Omega (\mathfrak Z , \mathfrak Z^*) = I_K\; ,
\label{resolid1}
\end{equation}
and the normalization condition:
\be
  \sum_{i=1}^M \langle \mathfrak Z; i\mid \mathfrak Z ; i\rangle = 1\; .
\label{VCS-normaliz}
\en

  The kernel $K$ has matrix elements
$$ K(\mathfrak Z^{*\prime} , \mathfrak Z )_{ij} =
     \chi^{i\dagger}K(\mathfrak Z^{*\prime} , \mathfrak Z )\chi^j\; . $$
But also, in view of (\ref{matrix-orthog}),
\bea
  \langle\mathfrak Z'; i \mid K(\mathfrak Z^* , \mathfrak Z )\mid\mathfrak Z; j\rangle
  & = &\int_\Omega \chi^{i\dagger}
  K(\mathfrak Z^{*\prime} , \mathfrak X )^* K(\mathfrak X^* , \mathfrak Z )\chi^j\;
  d\Omega (\mathfrak X, \mathfrak X^* )
     \nonumber\\
     & = & \chi^{i\dagger}K(\mathfrak Z^{*\prime} , \mathfrak Z )\chi^j =
     K(\mathfrak Z^{*\prime} , \mathfrak Z )_{ij} \; .
\label{kermatelem}
\ena

  Using (\ref{reker1}) the VCS can alternatively written as,
$$ \vert \mathfrak Z; i \rangle (\mathfrak Z^{*\prime })= \mathcal N(\mathfrak Z^* ,
\mathfrak Z )^{-\frac 12}
    \sum_k \frac {\mathfrak Z^{*\prime k}\; \mathfrak Z^k \chi^i}
                    {x_k!} = \mathcal N (\mathfrak Z^* , \mathfrak Z )^{-\frac 12}
  \sum_{j,k} \frac {\mathfrak Z^{*\prime k}\chi^j}{\sqrt{x_k!}}\cdot
                    \frac {\chi^{j\dagger} \mathfrak Z^k \chi^i}{\sqrt{x_k !}}, $$
so that,
\be
  \vert\mathfrak Z; i \rangle  = \mathcal N (\mathfrak Z^* , \mathfrak Z )^{-\frac 12}
  \sum_{j=i}^M \sum_{k=0}^N\bPhi^j_k \;
  \frac {\chi^{j\dagger}\mathfrak Z^k\chi^i}{\sqrt{x_k!}}\;.
\label{VCS2}
\end{equation}

   Let $\h$ be an $N$ dimensional (complex, separable) Hilbert space and let
$\{\phi_k\}_{k=0}^N$
be an orthonormal basis for it. Then the vectors $\chi^i\otimes \phi_k ,
\;\; 1=1,2, \ldots, M,\; k=0,1,2,
\ldots, N$, form an orthonormal basis of $\mathbb C^N\otimes \h$.
We make a unitary transformation,
$V: \hk \longrightarrow \mathbb C^N\otimes \h$, by the basis change $\bPhi^i_k \longmapsto
\chi^i\otimes\phi_k$. Under this map, the VCS $\vert \mathfrak Z; i\rangle$ transform to the vectors
\bea
  \vert \mathfrak Z , i\rangle^{\!\!\widetilde{\;\;\;}}& := & V\vert \mathfrak Z , i\rangle =
   \mathcal N (\mathfrak Z^* , \mathfrak Z )^{-\frac 12}
   \sum_{j=1}^M\sum_{k=0}^N\chi^j\otimes\phi_k \;
   \frac {\chi^{j\dagger}\mathfrak Z^k\chi^i}{\sqrt{x_k!}}\nonumber\\
   & = &
   \mathcal N (\mathfrak Z^* , \mathfrak Z )^{-\frac 12}\sum_{k=0}^N
   \frac {\mathfrak Z^k\chi^i}{\sqrt{x_k!}}
   \otimes \phi_k  \in \mathbb C^N\otimes\h\; ,
\label{VCS3}
\ena
which are exactly the VCS defined (over matrix domains) in \cite{aeg}. Also, in this
form the VCS resemble the non-linear coherent states (\ref{DCS})
more closely. The inverse
of the map $V$ is then easily seen to be given by,
\be
   (V^{-1}\bPhi )(\mathfrak Z^* ) = \sum_{i=1}^N \langle \mathfrak Z , i
   \vert\bPhi\rangle\chi^i \; , \qquad
      \bPhi \in \mathbb C^N\otimes\h \;.
\label{inv-isom}
\end{equation}

   To return to the discussion of the probability distribution $P(\lambda_1, \lambda_2,
\ldots , \lambda_M; n)$ in (\ref{comp-prob}), we first rewrite the VCS (\ref{VCS3}) explicitly
in matrix form as:
\be
\begin{split}
   & \quad \vert \mathfrak Z , i\rangle^{\!\!\widetilde{\;\;\;}}  = \\
  & \frac 1{\sqrt{M}}\sum_{k=0}^N
   U\begin{pmatrix} \sqrt{P(k,\lambda_1)}\;e^{-ik\theta_1} & 0 & \ldots & 0\\
                   0 &   \sqrt{P(k,\lambda_2)}\;e^{-ik\theta_2} & \ldots & 0\\
                   \vdots &\vdots & \ddots & \vdots\\
                   0&0&\ldots&   \sqrt{P(k,\lambda_M)}\;e^{-ik\theta_M}
              \end{pmatrix}U^*\chi^i\otimes \phi_k .
\end{split}
\label{explic-vcs}
\en
Again, let $\mathbb P_n =  \vert \phi_n\rangle\langle \phi_n\vert$ and define
\be
   \mathcal P (\mathfrak Z , \mathfrak Z^* ; n) = \text{Tr}_\h \left[\sum_{i=i}^M
   \vert \mathfrak Z , i\rangle^{\!\!\widetilde{\;\;}} \;^{\widetilde{\;\;}}\!\langle
   \mathfrak Z, i\vert \;\mathbb I_M\otimes\mathbb P_n\right]\; ,
\label{comp-prob-mat}
\en
where $\text{Tr}_\h$ denotes a partial trace in $\h$. Clearly,
$\mathcal P (\mathfrak Z , \mathfrak Z^* ; n ) $ is an $M\times M$ matrix and it is not
hard to see that
\be
  \mathcal P (\mathfrak Z , \mathfrak Z^* ; n )  =
   \frac 1M U\begin{pmatrix} P(n,\lambda_1) & 0 &\ldots & 0\\
                   0 &   P(n,\lambda_2) & \ldots & 0\\
                   \vdots &\vdots & \ddots & \vdots\\
                   0&0&\ldots &  P(n,\lambda_M)
              \end{pmatrix}U^*\; .
\label{comp-prob-mat2}
\en
Now taking the trace in $\mathbb C^M$ we immediately see that
\bea
   \text{Tr}_{\mathbb C^M} [\mathcal P (\mathfrak Z , \mathfrak Z^* ; n) ]
   & = & \text{Tr}_{\mathbb C^M\otimes\h} \left[\sum_{i=i}^M
   \vert \mathfrak Z , i\rangle^{\!\!\widetilde{\;\;}} \;^{\widetilde{\;\;}}\!\langle
   \mathfrak Z, i\vert \;\mathbb I_M\otimes\mathbb P_n\right]\nonumber\\
   & =&  P(\lambda_1, \lambda_2, \ldots , \lambda_M ;  n)\; ,
\label{comp-prob-mat3}
\ena
which should be compared with (\ref{prob-extrac}).
Finally, the determinant
\be
   \text{det}\left[M \mathcal P (\mathfrak Z , \mathfrak Z^* ; n ) \right]
    = P(n, \lambda_1)P(n, \lambda_2)\ldots P(n, \lambda_M)\; ,
\label{comp-prob-mat4}
\en
denotes the joint probability of getting $n$ ``successes'' from each distribution.

 Before leaving this topic of matrix valued distributions, let us point out that
more general situations than envisaged by (\ref{comp-prob}) can also be treated
using similar techniques. For example, instead of attaching the same weight,
$\dfrac 1M$, to each component $P(n, \lambda_i)$ of the mixture, we could also
attach different weights $\mu_i$ to them (with $\mu_i > 0$ for all $i$ and
$\sum_{i=1}^M \mu_i = 1$). Examples of this type will be dealt with in a future
publication, where we shall also allow the possibility of $M$ being infinite.

   To treat general joint probabilities of the type,
\be
   P(n_1, \lambda_1;\; n_2, \lambda_2; \; \ldots \; ; n_M , \lambda_M)
    = P(n_1, \lambda_1)P(n_2, \lambda_2)\ldots P(n_M, \lambda_M)\; ,
\label{multidim-prob}
\en
it is necessary to go to multidimensional coherent states. We intend to treat this in
greater detail in a future publication, but here we briefly indicate the main idea.
Consider again a discrete distribution $P(n, \lambda )$ of the type
(\ref{CS-assoc-discr-distrib}), i.e., such that it has associated coherent states of the
type (\ref{back-constr}). These coherent states $\vert z \rangle$ are defined on a
Hilbert space $\h$. Let $\h^M = \h\otimes \h\otimes\ldots\otimes\h$ be the $M$-fold
tensor product of $\h$ with itself. On $\h^M$ we define the vectors,
\bea
  \vert z_1, z_2 , \ldots , z_M\rangle & = & \vert z_1 \rangle\vert z_2 \rangle
    \ldots \vert z_M \rangle\nonumber\\
    & = & \sum_{n_1=0,\; n_2 = 0,\; \ldots , \; n_M=0}^N
    [P(n_1, \lambda_1;\; n_2, \lambda_2; \; \ldots \; ; n_M , \lambda_M)]^{\frac 12}
    \nonumber\\
    &\times& e^{i(n_1\theta_1 + n_2\theta_2 +\; \ldots \;+ n_M\theta_M )} \phi_{n_1 ,\; n_2,\;
    \ldots, \; n_M}\; ,
    \label{multidim-cs}
\ena
where the vectors
$$ \phi_{n_1 ,\; n_2,\; \ldots, \; n_M} =
   \phi_{n_1}\otimes\phi_{n_2}\otimes\ldots\otimes\phi_{n_M}\; ,
\qquad 0\leq n_1 ,\; n_2,\;
    \ldots ,\;  n_M \leq N\; , $$
form an orthonormal basis for $\h^M$. We call the vectors (\ref{multidim-cs})
{\em multidimensional coherent states\/.} Such coherent states have been studied in
different contexts before (see, for example, \cite{novgaz}). These vectors are normalized,
$$ \langle z_1, z_2 , \ldots , z_M \mid z_1, z_2 , \ldots , z_M \rangle =1, $$
and they satisfy the resolution of the identity (compare with (\ref{resolid7})),
$$
  \int_{\mathcal D^M} \vert z_1, z_2 , \ldots , z_M \rangle\langle z_1, z_2 , \ldots , z_M\vert\;
  \prod_{i=1}^M \mathcal N(\lambda_i )\; d\overline{\varrho} (\lambda_i ) \; d\theta_i
  = I_{\h^M}\;, $$
where $\mathcal D^M = \mathcal D\times\mathcal D\times\ldots\times\mathcal D$  is
the $M$-fold cartesian product of the domain $\mathcal D = \{ \sqrt{\lambda}\; e^{i\theta}
\in \mathbb C \mid \lambda \in [0, L), \;\; \theta \in [0, 2\pi ) \}$ over which the coherent
states $\vert z \rangle$ in (\ref{back-constr}) are defined.

   Once again these coherent states appear as ``generalized square-roots'' of the
joint probability distribution $P(n_1, \lambda_1;\; n_2, \lambda_2; \; \ldots \; ; n_M ,
\lambda_M),
\;\; 0\leq n_1 ,\; n_2,\; \ldots \; n_M \leq N$, and just as in (\ref{prob-extrac}),
\bea
  P(n_1, \lambda_1;\; n_2, \lambda_2; \; \ldots \; ; n_M , \lambda_M) & = &
  \text{Tr} \;[\vert z_1, z_2, \ldots , z_M\rangle \langle z_1, z_2, \ldots , z_M \vert
   \;\mathbb P_{n_1 ,\; n_2,\; \ldots ,\; n_M}]\nonumber\\
   & =& \vert\langle\phi_{n_1 ,\; n_2,\; \ldots \; n_M}
   \mid z_1, z_2, \ldots , z_M \rangle\vert^2 \; ,
\label{multidim-prob-ext}
\ena
with
$$
  \mathbb P_{n_1 ,\; n_2,\; \ldots ,\; n_M} = \vert \phi_{n_1 ,\; n_2,\; \ldots , \; n_M}
  \rangle\langle \phi_{n_1 ,\; n_2,\; \ldots ,\; n_M}\vert\; . $$

Recently (see \cite{gagaque}), this formalism has been applied to the construction of vector
coherent states  for the quantum motion of a particle in an infinite square well,
enabling one to define in an unambiguous way the momentum operator. The construction
in \cite{gagaque} is based
on  Gaussian probability distributions but it can be carried out using  a large
class of distributions.

\section{Conclusion}
 As mentioned in the Introduction, the relationship between coherent states and
statistical distributions has been studied before. We have tried to demonstrate here
the deeper connection between such distributions, both continuous and discrete, and
reproducing kernel Hilbert spaces, in so far as the latter
are the carriers of generalized
coherent states. Moreover, taking this point of view, it has been possible to connect
vector coherent states to mixtures of
probability distributions  and  multi-dimensional
coherent states to joint probability distributions. The posterior distribution,
appearing  on the
parameter space of a discrete distribution, is clearly seen to be a consequence of the
resolution of the identity satisfied by the coherent states. Again this has been
noticed earlier, but here we are able to put it in a more general context.

  The one
intriguing question that arises from the general discussion is the following: as has
been demonstrated, a discrete statistical distribution, or a family of discretely
parametrized continuous distributions, satisfying certain technical
conditions, lead to the existence of coherent states on an associated Hilbert space.
These coherent states, in turn, can be shown to lead to quantum probabilities,
embodied in a positive operator valued measure, on the
parameter space. The nature of classical (commutative) and quantum (non-commutative)
probability are intrinsically different, yet it seems to be possible to make a smooth
transition from one to the other. This is reminiscent of the process of quantization, i.e.,
the passage from a classical mechanical system to its quantum counterpart,
and in particular, coherent state quantization (see, for example, \cite{alieng}
for a review of the theory of quantization and
\cite{gagaque,gapi,gmm,gaga,gahulare1} for a series of examples).
So one might ask
the question as to whether the
procedure described above could be considered as constituting a quantization of the underlying
classical probability theory. In this connection it would also be interesting to study more
closely the duality appearing between the discrete and continuous distributions incorporated
in the coherent states and the analogous duality familiar from Bayesian statistics.

\section*{Appendix: Some elements of Bayesian inference}

In this Appendix we put together some notions from Bayesian statistical
inference that have been used in this paper. Some relevant references are
\cite{kojo,hels,holev,per}
\subsection*{Event space background}    The context is the setup and subsequent
performance of an experiment where there is a random component to the results and
where the set $U$ of possible results is known. In the field of statistics, the
experiment is called a ``random experiment''.  Events are identified with measurable
subsets of $U$. That is, we say that event $E$ has occurred if the observed
result $u_{\mathrm{obs}}$ is in the subset $E$. One ``experiment'', of course,
could be an amalgam of a whole set of sub-experiments, sometimes
called ``trials''.

\subsection*{Conditional probabilities}
    Let $P(E \mid B)$ designate the conditional probability that event $E$ occurs
    given that event $B$ has occurred. Then
\begin{equation*}
 P(E \mid B) =  \frac {P(E  \cap B)}{P(B)}\, ,
\end{equation*}
where the numerator stands for the joint probability of  occurrence of events $E$
and $B$ and the denominator is the unconditional probability of occurrence of event
$B$ (to ensure normalization). Consider the conditional probability the other way around $P(B \mid E)$.
\begin{equation*}
 P(B \mid E) =  \frac{P(E \cap  B)}{P(E)}\, .
\end{equation*}
Suppose that we do not know the joint probability and in fact we only know the first
conditional probability  $P(E\mid B)$ and the two unconditional probabilities, then we can write
\begin{equation*}
 P(B \mid E) =  \frac {P(E\mid  B)\, P(B)}{P(E)}\, .
\end{equation*}
The probability $P(B \mid E)$ is called the {\em posterior conditional
probability} for $B$ given $E$ and $P(B)$ is called the {\em prior probability} of $B$.
Sometimes we compute several of these posterior probabilities in the cases where
the set of events $\{B_1, B_2,\dotsc ,B_n\}$ is a partition of $U$ and the
events $B_i$ are in the nature of possible causal hypotheses for the subsequent
occurrence of event $E$.  Suppose that we know the conditional probabilities
$P(E\mid B_i)$ and the unconditional (prior) probabilities $P(B_i)$ for each
$B_i$. Then one chooses a likely hypothesis by computing each of the posterior
probabilities.

\subsection*{The case of a continuous family of discrete probability distributions}

Consider the performance of a classical experiment in
which the outcome has a random component within the following context.
Let $n = 0,1,2,\ldots , N$  index the (discrete) set of possible outcomes
of the experiment, where $N$ is a positive integer or $\infty$. For
real parameter $\lambda \in \Lambda$,
let $P(n, \lambda)$ be a family of classical discrete probability distributions
indexed by $\lambda$, which serves as a stochastic model for the experiment. We
suppose that $\lambda$ is unknown and the object of the experiment is to obtain
data with which to infer a probability distribution on the parameter space
$\Lambda$.
After performance of the experiment, let $k$ indicate the observed outcome.
Then construct a conditional probability density function $f$ for $\lambda$,
given $k$, in the form:
$$
 f(\lambda , k)  =   \frac {P(k, \lambda)\Pi (\lambda )}
              {\int_{\Lambda} P(k, \lambda' )\Pi (\lambda' )\; d\lambda' }\;,$$
where $\Pi (\lambda )$ is an unconditional probability measure on the parameter
space $\Lambda$, arbitrary, subject to the integrability of the denominator.
The measure $\Pi$ is called the {\em prior measure} on $\Lambda$  and the
conditional
probability density function $f$ is called the density function of the
{\em posterior probability distribution} on $\Lambda$.

\bigskip

\noindent{\em Example :}  Toss a coin $N$ times observing $n$,
the number of occurrences of
heads. Let the parameter $p$ be the probability of obtaining heads on one toss.
Supposing that $p$ is unknown, the object is to use the outcome of the
experiment to obtain a probability distribution on the parameter space
$(0,1)$.  The stochastic model is the binomial family,
$$ P(n,p)  = \frac {N!}{(N-n)! n!} p^n (1-p)^{(N-n)}\; , \quad \text{for} \quad
n = 0,1,2,...,N,$$
where $N$ is a positive integer.
After the performance of the experiment, having obtained $k$ heads,
with choice of prior measure $\Pi (p)$, the posterior distribution on
$(0,1)$ is given by the conditional probability density function,
$$
  f(p,k) =   \frac {p^k (1- p)^{(N-k)} \Pi (p)}
      {\int_0^1 p'^k (1- p')^{(N-k)} \Pi (p')\; dp'}\; . $$

\section*{Acknowledgements}

The work of one of the authors (STA) was partially supported
through grants from the Natural Sciences and Engineering Research
Council (NSERC), Canada and the Fonds qu\'eb\'ecois de la recherche
sur la nature et les technologies (FQRNT), Qu\'ebec. The authors would also
like to acknowledge useful discussions with M. Wang.

\end{document}